\documentclass[conference]{IEEEtran}
\IEEEoverridecommandlockouts
\usepackage{cite}
\usepackage{amsmath,amssymb,amsfonts}
\usepackage{algorithmic, algorithm}
\usepackage{graphicx}
\usepackage{textcomp}
\usepackage{xcolor}
\usepackage{tcolorbox}
\usepackage{url}
\usepackage{booktabs}
\usepackage{hyperref,cleveref}
\usepackage{setspace}
\usepackage{balance}
\def\BibTeX{{\rm B\kern-.05em{\sc i\kern-.025em b}\kern-.08em
    T\kern-.1667em\lower.7ex\hbox{E}\kern-.125emX}}
    
\begin{document}


\title{BotRGCN: Twitter Bot Detection with Relational Graph Convolutional Networks
}

\author{\IEEEauthorblockN{Shangbin Feng, Herun Wan, Ningnan Wang, Minnan Luo}
\IEEEauthorblockA{\textit{School of Electronic and Information Engineering} \\
\textit{Xi'an Jiaotong University}\\
Xi'an, China \\
\{wind\_binteng,wanherun,mrwangyou\}@stu.xjtu.edu.cn,minnluo@xjtu.edu.cn}}


\maketitle

\begin{abstract}
Twitter bot detection is an important and challenging task. Existing bot detection measures fail to address the challenge of community and disguise, falling short of detecting bots that disguise as genuine users and attack collectively. To address these two challenges of Twitter bot detection, we propose BotRGCN, which is short for Bot detection with Relational Graph Convolutional Networks. BotRGCN addresses the challenge of community by constructing a heterogeneous graph from follow relationships and applies relational graph convolutional networks. Apart from that, BotRGCN makes use of multi-modal user semantic and property information to avoid feature engineering and augment its ability to capture bots with diversified disguise. Extensive experiments demonstrate that BotRGCN outperforms competitive baselines on a comprehensive benchmark TwiBot-20 which provides follow relationships.
\end{abstract}

\section{Introduction}
Twitter is a thriving social media platform with millions of daily active users. Besides being home to genuine users, Twitter is also home to automated programs, also known as Twitter bots. These bots are operated to induce undesirable social effects such as extreme propaganda~\cite{berger2015isis} and election interference~\cite{10.1145/3308560.3316486, DBLP:journals/corr/Ferrara17aa}. That being said, there is an urgent need for robust Twitter bot detectors.

Existing methods generally fall into two categories: feature engineering and deep learning. For feature engineering based bot detectors, user features such as tweet features~\cite{miller2014twitter}, user property features~\cite{d2015real} and features extracted from neighborhood information~\cite{yang2013empirical} were adopted with traditional classifiers. For deep bot detection models, recurrent neural networks~\cite{wei2019twitter,kudugunta2018deep} and generative adversarial networks~\cite{stanton2019gans} were adopted.

Despite early successes, the ever-changing social media has brought two new challenges to the task of Twitter bot detection: disguise and community. The challenge of disguise demands bot detectors to capture malicious bots even when they are designed to resemble genuine users. For example, Cresci \textit{et al.}\cite{10.1145/3409116} spotted bots that use stolen names and profile pictures and intersperse few malicious messages with many neutral ones. Apart from that, the challenge of community demands bot detectors to successfully capture Twitter bots that seem genuine individually but act in groups to pursue malicious goals. For example, Cresci \textit{et al.} \cite{cresci2017paradigm} identified a group of bots and their collective action towards influencing the mayoral election of Rome in 2014.

In light of the two challenges of Twitter bot detection, we propose a novel framework BotRGCN (\textbf{Bot} detection with \textbf{R}elational \textbf{G}raph \textbf{C}onvolutional \textbf{N}etworks). Specifically, BotRGCN addresses the challenge of disguise by leveraging all available numerical and categorical user property items and encoding user tweets with pre-trained language models. BotRGCN addresses the challenge of community by constructing a heterogeneous graph from the Twitter network and apply relational graph convolutional networks. 

\begin{figure}[t]
\centerline{\includegraphics[width=\linewidth]{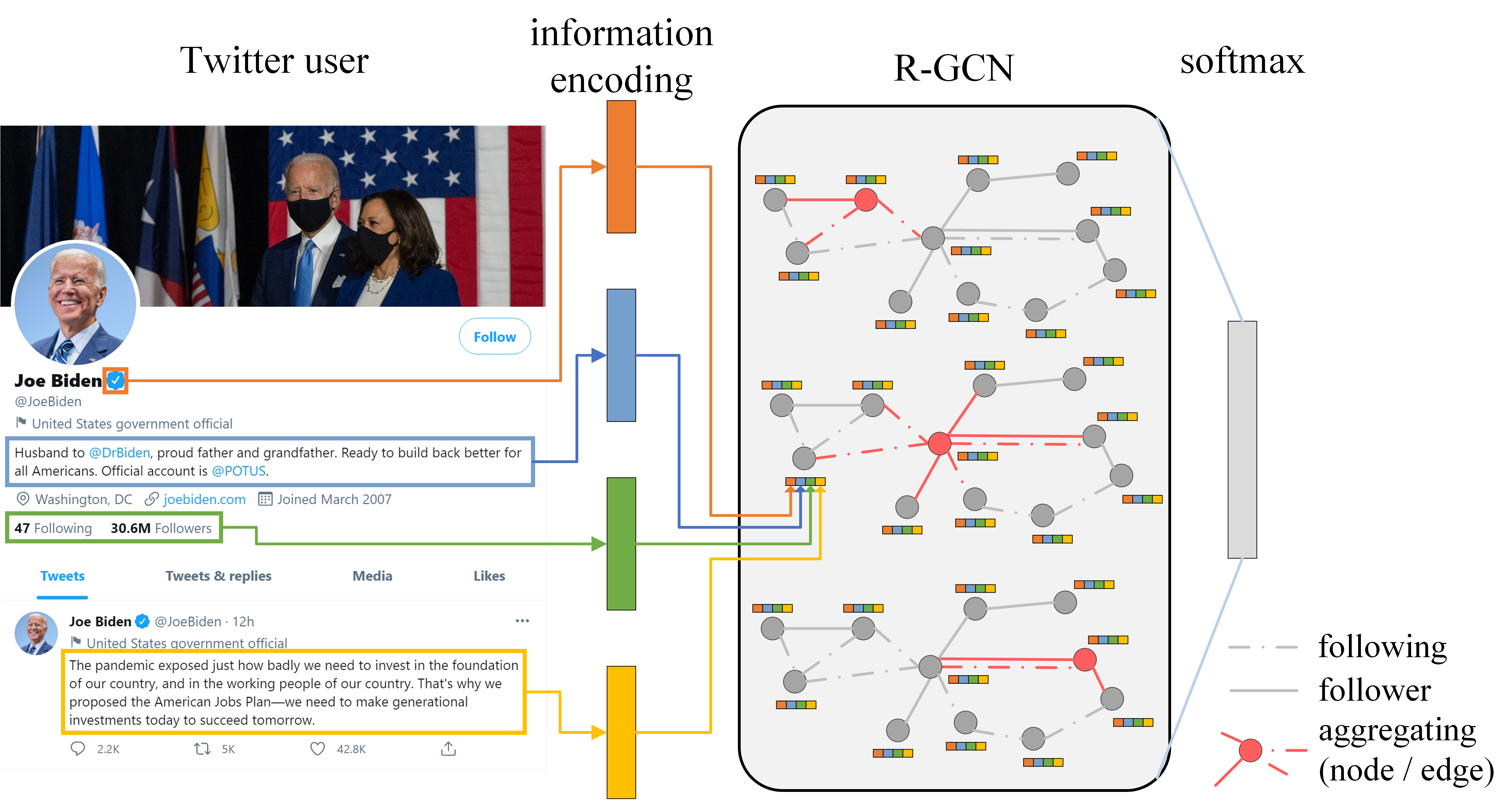}}
\caption{Overview of BotRGCN. Orange, blue, green and yellow modules denote categorical metadata, user description, numerical metadata and tweets.}
\label{fig:overview}
\end{figure}

\section{BotRGCN Methodology}
\label{sec:methodology}

\subsection{Problem Definition}
\label{sec:problem_definition}
Let $B = \{b_i\}^L_{i=1}$ denotes a user's description with $L$ words. Let $T = \{t_i\}_{i=1}^{M}$ be a user's $M$ tweets and each tweet $t_i = \{w_1^i, \cdot \cdot \cdot, w_{Q_i}^i\}$ contains $Q_i$ words. Let $P = \{P^{num},P^{cat}\}$ be a user's numerical and categorical user property set. Let $N = \{N^f, N^t\}$ be a user's neighborhood information, where $N^f = \{N_1^f,\cdot \cdot \cdot, N^f_u\}$ denotes user's followings and $N^t = \{N_1^t, \cdot \cdot \cdot, N_v^t\}$ denotes user's followers. The task of Twitter bot detection is to identify bots among users with the help of user information $B$, $T$, $P$ and $N$.

\subsection{User Feature Encoding}
\label{subsec:method_feature_encoding}
BotRGCN is designed to address the challenge of disguise by leveraging multi-modal user information, which leaves bot operators no venue to achieve malicious goals. Specifically, BotRGCN jointly encodes user semantic information of description and tweets as well as both numerical and categorical user property information.

\begin{table}[t]
\caption{Numerical user properties adopted in BotRGCN.}
\label{tab:numerical_feature}
\begin{center}
\begin{tabular}{cc}
\toprule 
Feature Name & Description \\ \midrule
\#followers & number of followers \\ 
\#followings & number of followings  \\ 
\#favorites & number of likes \\  
\#statuses & number of statuses \\ 
active\_days & number of active days \\ 
screen\_name\_length & screen name character count \\ 
\bottomrule
\end{tabular}
\end{center}
\end{table}

\paragraph{Overall user feature vector} We encode user description, tweets, numerical and categorical properties and concatenate them to serve as user features:
\begin{equation}
\label{equ:overall_feature}
    r = [r_b;r_t;r_p^{num};r_p^{cat}] \in \mathbb{R}^{D \times 1}
\end{equation}
where $D$ is the user embedding dimension. We present BotRGCN's strategy of encoding user desciption, tweets, numerical and categorical property items in the following.

\paragraph{Feature set 1: user description} We adopt pre-trained RoBERTa\cite{liu2019roberta} to encode user descriptions. We firstly transform words in user description with RoBERTa:
\begin{equation}
    \bar{b} = RoBERTa(\{b_i\}_{i=1}^L),\ \ \ \bar{b} \in \mathbb{R}^{D_s \times 1}
\end{equation}
where $\bar{b}$ denotes representation of user description and $D_s$ is the RoBERTa embedding dimension. We then derive representation vectors for user's description:
\begin{equation}
\label{equ:S1end}
    r_b = \phi(W_B \cdot \bar{b} + b_B),\ \ \ r_b \in \mathbb{R}^{D/4 \times 1}
\end{equation}
where $W_B$ and $b_B$ are learnable parameters, $\phi$ is the activation function and $D$ is the embedding dimension of Twitter users. We adopt leaky-relu as $\phi$ for the rest of the paper.
\paragraph{Feature set 2: user tweets} We use RoBERTa to similarly encode user tweets. We average the representation of all tweets to obtain representation of user tweets $r_t$.

\paragraph{Feature set 3: user numerical properties} BotRGCN leaves handling of user property items to MLPs and graph neural networks. Specifically, we adopt numerical featrues that are directly available from the Twitter API without feature engineering and present them in \Cref{tab:numerical_feature}. Specifically, we conduct z-score normalization and obtain representation of user numerical features $r_p^{num}$ with a fully connected layer.

\paragraph{Feature set 4: user categorical properties} Similar to user numerical properties, we avoid feature engineering and apply MLPs and graph neural networks to encode them. We leverage directly available user categorical features from the Twitter API and they are presented in \Cref{tab:categorical_feature}. Specifically, we adopt one-hot encoding, concatenate and transform them with a fully connected layer and leaky-relu to derive representation for user's categorical features $r_p^{cat}$.

\begin{table}[t]
\caption{Categorical user properties adopted in BotRGCN.}
\label{tab:categorical_feature}
\begin{center}
\resizebox{\linewidth}{!}{\begin{tabular}{cc}
\toprule
Feature Name & Description \\ \midrule
protected &  protected or not \\ 
geo\_enabled &  enable geo-location or not \\ 
verified &  verified or not \\ 
contributors\_enabled & enable contributors or not \\ 
is\_translator &  translator or not \\ 
is\_translation\_enabled & translation or not  \\ 
profile\_background\_tile & the background tile \\ 
profile\_user\_background\_image & have background image or not \\ 
has\_extended\_profile & have extended profile or not \\ 
default\_profile & the default profile \\ 
default\_profile\_image & the default profile image \\ 
\bottomrule
\end{tabular}}
\end{center}
\end{table}














\subsection{GNNs Architecture}
\label{subsec:method_BotRGCN}
BotRGCN is designed to address the challenge of community by leveraging user follow relationship and the dense graph structure it forms. Specifically, BotRGCN constructs a heterogeneous graph from the Twitter network and apply relational graph convolutional networks to learn user representations.

\paragraph{Graph construction} BotRGCN treats Twitter users as nodes. Given that following and being followed signal different information, BotRGCN leverages two types of edges, $R = \{r_1,r_2\} = \{"following", "follower"\}$. We denote user $u$'s following and follower neighborhood as $N_{r_1}(u) = N^f(u)$ and $N_{r_2}(u) = N^t(u)$. By defining two sets of relational neighborhood for each Twitter user, BotRGCN constructs a heterogeneous graph that reflects the interactions between Twitter users. BotRGCN could incorporate more relation types between users if supported by the data set.

\paragraph{BotRGCN architecture}
We apply R-GCNs\cite{R-GCN} to the heterogeneous graph and learn user representations. Specifically, we firstly transform user features to derive the initial hidden vectors for nodes in the graph:
\begin{equation}
\label{equ:GNNbegin}
    x_i^{(0)} = \phi(W_1 \cdot r_i + b_1),\ \ \ x_i^{(0)} \in \mathbb{R}^{D \times 1}
\end{equation}
where $W_1$ and $b_1$ are learnable parameters. We then apply the $l$-th R-GCN layers:
\begin{equation}
    x_i^{(l+1)} = \Theta_{self} \cdot x_i^{(l)} + \sum_{r\in R} \sum_{j\in N_r(i)} \frac{1}{|N_r(i)|} \Theta_r \cdot x_j^{(l)}, x_i^{(l+1)} \in \mathbb{R}^{D\times 1}
\end{equation}
where $\Theta$ is the projection matrix. After $L$ layers of R-GCN, we transform the user representation with MLP:
\begin{equation}
\label{equ:GNNend}
    h_i = \phi(W_2 \cdot x_i^{(L)} + b_2),\ \ \ h_i \in \mathbb{R}^{D \times 1}
\end{equation}

\begin{table}[t]
\caption{Bot detection performance on TwiBot-20 benchmark.}
\label{tab:big}
\begin{center}
\begin{tabular}{cccc}
\toprule
Method &\ Accuracy\ &\ F1-score\ & MCC\cite{matthews1975comparison}\\ \midrule
Lee \textit{et al.} \cite{lee2011seven} & 0.7456 & 0.7823 & 0.4879 \\ 
Yang \textit{et al.} \cite{yang2020scalable}& 0.8191 & 0.8546 & 0.6643 \\ 
Kudugunta \textit{et al.} \cite{kudugunta2018deep}& 0.8174 & 0.7517 & 0.6710 \\ 
Wei \textit{et al.} \cite{wei2019twitter}& 0.7126 & 0.7533 & 0.4193 \\ 
Miller \textit{et al.} \cite{miller2014twitter}& 0.4801 & 0.6266 & -0.1372 \\ 
Cresci \textit{et al.} \cite{cresci2016dna}& 0.4793 & 0.1072 & 0.0839 \\ 
Botometer \cite{davis2016botornot}& 0.5584 & 0.4892 & 0.1558 \\ 
Alhosseini \textit{et al.} \cite{ali2019detect}& 0.6813 & 0.7318 & 0.3543 \\ 
SATAR~\cite{feng2021satar} & 0.8412 & 0.8642 & 0.6863 \\
\textbf{BotRGCN} & \textbf{0.8462} & \textbf{0.8707} & \textbf{0.7021} \\ 

\bottomrule
\end{tabular}
\end{center}
\end{table}












\noindent where $W_2$ and $b_2$ are learnable parameters and $h_i$ is the representation for user $i$. 

\subsection{Learning and Optimization}
\label{subsec:method_learning}
We apply a softmax layer to conduct Twitter bot detection based on user representations derived from R-GCN :
\begin{equation}
\label{equ:lossbegin}
    \hat{y_i} = softmax(W_O \cdot h_i + b_O)
\end{equation}
where $W_O$ and $b_O$ are learnable parameters.

The loss function of BotRGCN is constructed as follows:
\begin{equation}
\label{equ:lossend}
    L = -\sum_{i\in Y}[y_ilog(\hat{y_i}) + (1-y_i)log(1-\hat{y_i})] + \lambda \sum_{w\in \theta}w^2
\end{equation}
where $Y$ denotes annotated users, $y_i$ is the ground-truth label and $\theta$ are all learnable parameters in the BotRGCN framework. 

\section{Experiments}
\label{sec:experiments}

\subsection{Experiment Settings}

\paragraph{Dataset} TwiBot-20~\cite{feng2021twibot} is a publicly available Twitter bot detection dataset that provides follow relationship between users to support BotRGCN. We adopt TwiBot-20 and follow the partition of train, validation and test set in the original benchmark. We obtain a heterogeneous graph with 229,580 nodes and 227,979 edges from the data set.

\paragraph{Baseline methods} We compare BotRGCN with the following baselines:

\begin{itemize}

\item Lee \textit{et al.}\cite{lee2011seven}: Lee \textit{et al.} use random forest with several user features, e.g. the longevity of the account.

\item Yang \textit{et al.}\cite{yang2020scalable}: Yang \textit{et al.} use random forest with minimal account metadata.

\item Kudugunta \textit{et al.}\cite{kudugunta2018deep}: Kudugunta \textit{et al.} propose an architecture that uses both tweet content and the metadata. 

\item Wei \textit{et al.}\cite{wei2019twitter}: Wei \textit{et al.} use word embeddings and a three-layer BiLSTM for bot detection.

\item Miller \textit{et al.}\cite{miller2014twitter}: Miller \textit{et al.} extract 107 features from a user's tweet and property information. It conducts bot detection as anomaly detection.

\item Cresci \textit{et al.}\cite{cresci2016dna}: Cresci \textit{et al.} use strings to represent the sequence of a user's online actions. It identifies bot groups by analyzing longest common substrings.

\item Botometer\cite{davis2016botornot}: Botometer 
is a publicly available service that leverages more than one thousand features.

\item Alhosseini \textit{et al.}\cite{ali2019detect}: Alhosseini \textit{et al.} use graph convolutional networks to detect Twitter bots.

\item SATAR~\cite{feng2021satar} constructs a self-supervised task for Twitter user representation learning and applies it to the task of bot detection with fine-tuning.
\end{itemize}


\vspace{-5pt}
\subsection{Bot Detection Performance}
\vspace{-2pt}
\Cref{tab:big} presents bot detection performance on TwiBot-20. It is demonstrated that BotRGCN achieves state-of-the-art performance among all methods, which demonstrates that BotRGCN is generally effective in the task of Twitter bot detection. Besides, BotRGCN outperforms baselines that also leverage user follow relationship such as Alhosseini \textit{et al.} \cite{ali2019detect} and SATAR~\cite{feng2021satar}, which shows that BotRGCN better utilizes follow relationships that put users into their social context.

\begin{figure}[t]
\centerline{\includegraphics[width=\linewidth]{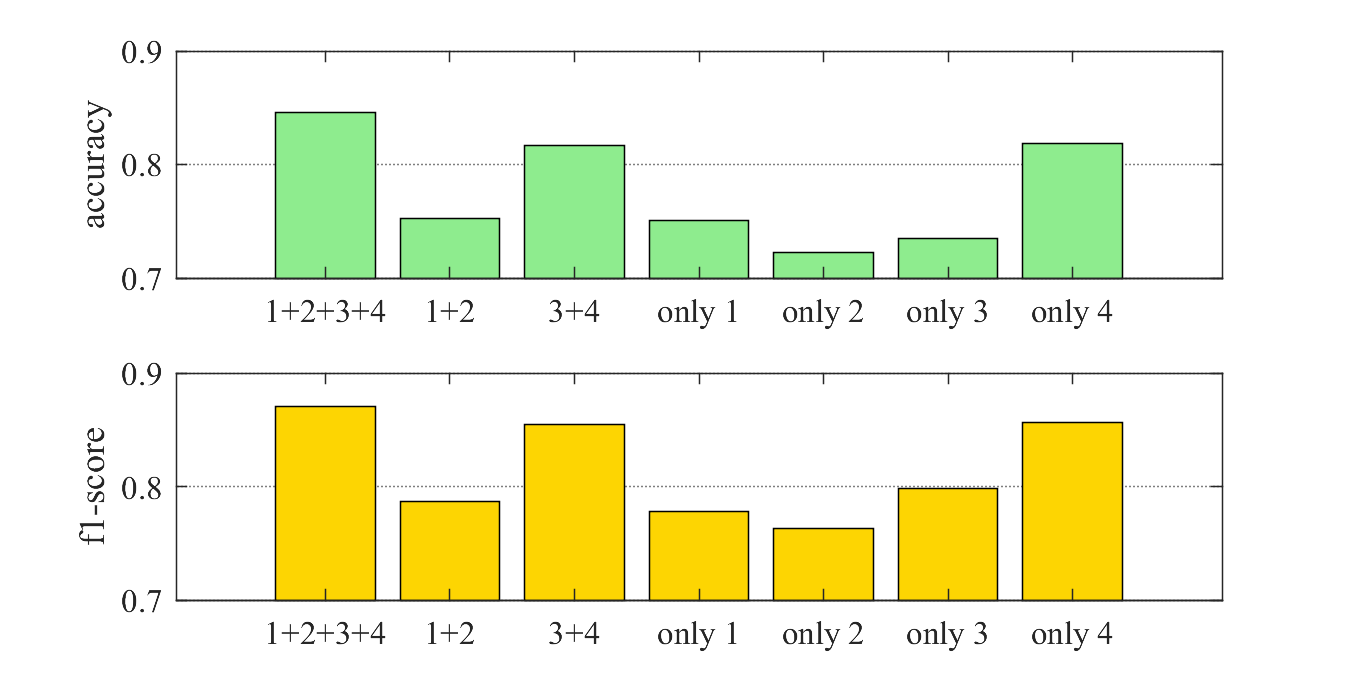}}
\caption{BotRGCN performance with different user feature sets.}
\label{fig:feature_study}
\end{figure}

\begin{figure}[t]
\centerline{\includegraphics[width=\linewidth]{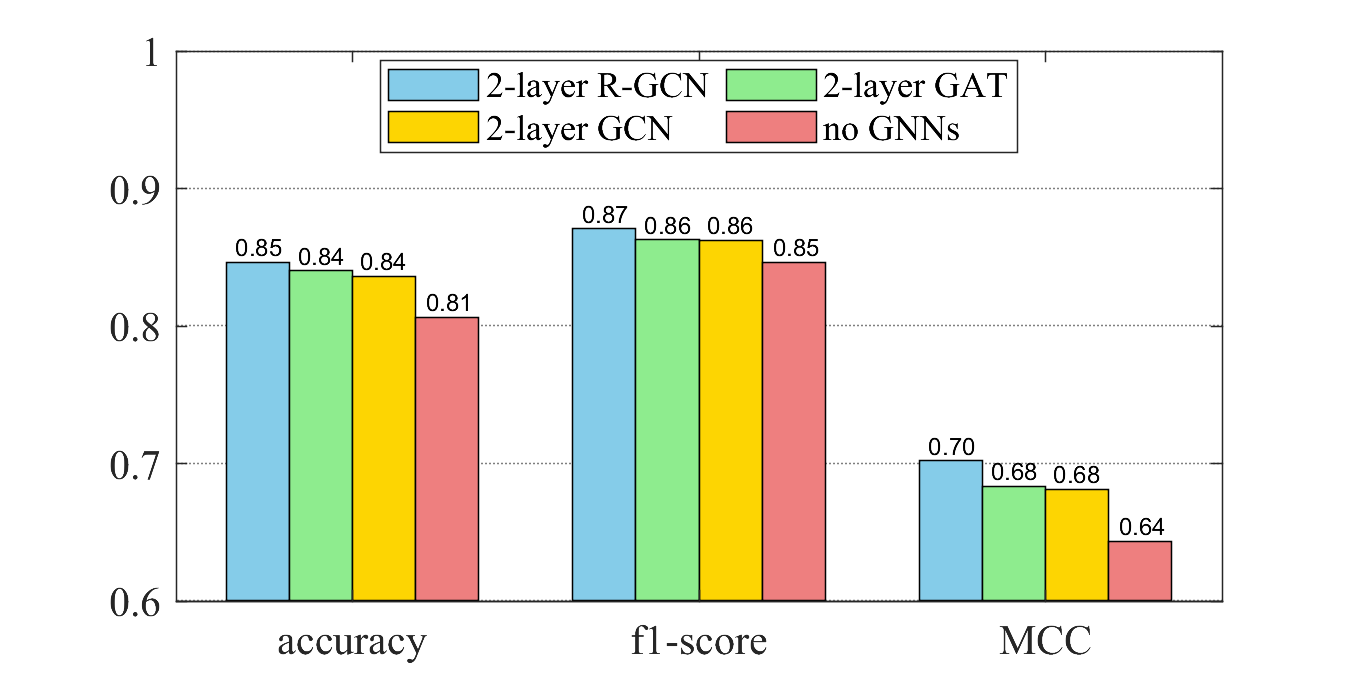}}
\caption{BotRGCN performance with different types of graph neural networks.}
\label{fig:GNN_type_study}
\end{figure}

\subsection{User Feature Study}
To prove that jointly encoding multi-modal user information is necessary for robust bot detectors, we conduct ablation study to train BotRGCN with reduced feature sets and present the results in \Cref{fig:feature_study}. It is demonstrated that every aspect of user information is essential in BotRGCN's performance, while user categorical properties contribute most to its performance.

\subsection{GNN Study}
To examine the necessity of R-GCN and the possibility of using other graph neural networks on a homogeneous graph, we substitute R-GCN in BotRGCN with GAT\cite{GAT}, GCN\cite{GCN} and MLP and present the results in \Cref{fig:GNN_type_study}. It indicates that our choice of R-GCN contributes to BotRGCN's performance.

\begin{figure}[t]
\centerline{\includegraphics[width=\linewidth]{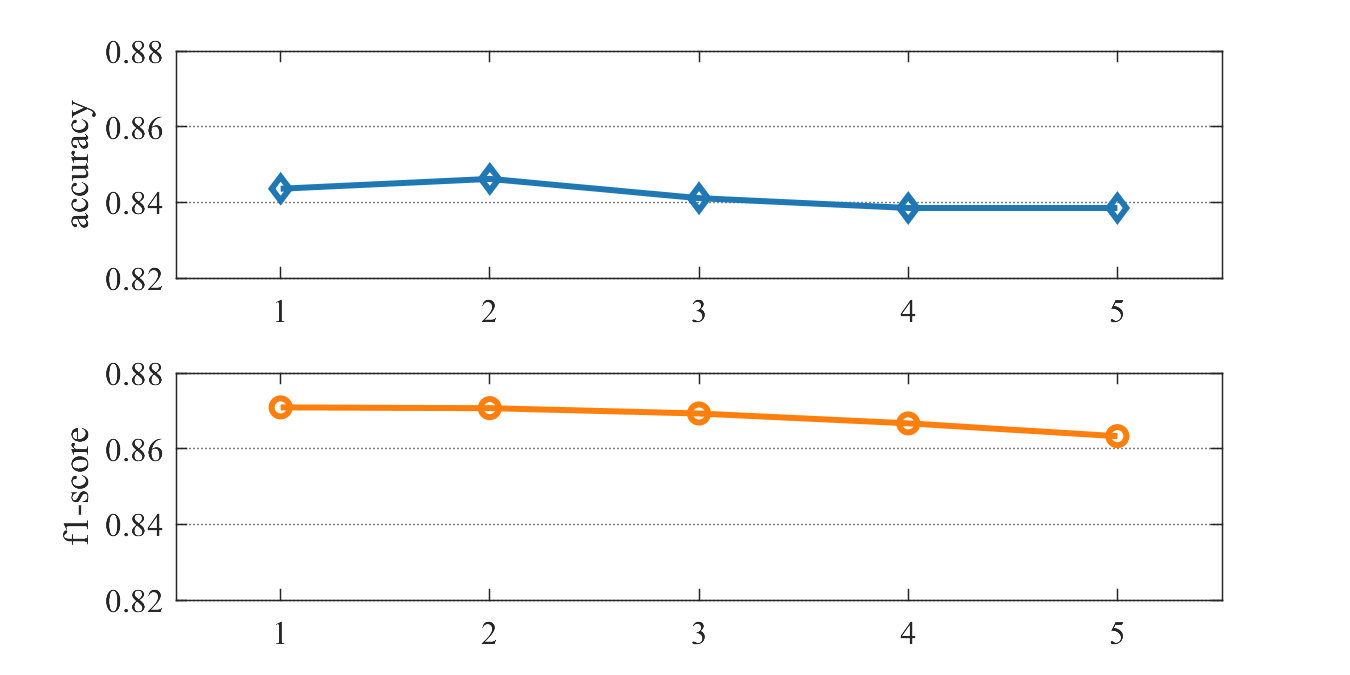}}
\caption{BotRGCN performance with different number of R-GCN layers.}
\label{fig:GNN_layer_study}
\end{figure}

We further explore different amount of R-GCN layers and its effect on the overall bot detection performance. Result in \Cref{fig:GNN_layer_study} shows that BotRGCN with 2 layers of R-GCN could result in better bot detection performance with fewer learnable parameters and less training complexity.

\section{Conclusion}
\label{sec:conclusion}
Social media bot detection is attracting growing attention. We proposed BotRGCN, an end-to-end bot detection framework that jointly encodes multi-modal user information, construct a heterogeneous graph to represent real-world Twitter and apply relational graph convolutional networks. BotRGCN is designed to tackle the challenges of bot disguise and bot communities. We conduct extensive experiments to demonstrate the efficacy of BotRGCN in comparison to state-of-the-art baseline methods. Further explorations proved that BotRGCN's user information encoding strategy and its graph learning approach are essential to the model's performance.

\bibliographystyle{IEEEtran}
\balance
\bibliography{sample-base}

\end{document}